\documentclass[
  journal=pasa,
  manuscript=research-paper,
  year=2025,
  volume=YY,
]{cup-journal}

\usepackage{amsmath}
\usepackage{amssymb,microtype,siunitx,booktabs}

\usepackage{mathrsfs}
\usepackage{hyperref}
\usepackage{orcidlink}
\usepackage{lineno}
\usepackage{xcolor}

\hypersetup{
    colorlinks=true,
    linkcolor=blue,
    filecolor=blue,
    urlcolor=blue,
    citecolor=blue,
    pdfborder={0 0 0}
}

\newcommand{\ash}{{$\textsc{Ashvini}$}}

\newcommand{\Zg}{{M_{\rm Z,g}}}
\newcommand{\Zstar}{M_{\rm Z,\star}}

\newcommand{\Zsun}{\text{Z}_\odot}
\newcommand{\Msun}{\text{M}_\odot}

\newcommand{\Mg}{{M_{\rm g}}}
\newcommand{\Mstar}{M_{\rm \star}}
\newcommand{\Mdust}{M_{\rm d}}
\newcommand{\Mhalo}{M_{\rm h}}

\defcitealias{Menon2024}{MP24}
\received {dd Mmm YYYY}
\revised  {dd Mmm YYYY}
\accepted {dd Mmm YYYY}
\published {dd Mmm YYYY}

\title[On bursty star formation during cosmological reionization --- influence on the metal and dust content of low-mass galaxies]{On bursty star formation during cosmological reionization --- influence on the metal and dust content of low-mass galaxies}

\author{Anand Menon\orcidlink{0009-0004-2816-9913}}
\affiliation{International Centre for Radio Astronomy Research, The University of Western Australia, 35 Stirling Highway, Crawley, Western Australia 6009, Australia}
\alsoaffiliation{The Oskar Klein Centre and Department of Physics, Stockholm University, AlbaNova University Centre, SE 106 91 Stockholm,
Sweden}

\author{Sreedhar Balu\orcidlink{0000-0002-5281-5151}}
\affiliation{Facultad de Físicas, Multidisciplinary Unit for Energy Science, Universidad de Sevilla, 41012, Seville, Spain}

\author{Chris Power\orcidlink{0000-0002-4003-0904}}
\affiliation{International Centre for Radio Astronomy Research, The University of Western Australia, 35 Stirling Highway, Crawley, Western Australia 6009, Australia}
\alsoaffiliation{ARC Centre of Excellence for All Sky Astrophysics in 3 Dimensions (ASTRO 3D)}

\email[Anand Menon]{anand.menon@fysik.su.se}

\keywords{galaxies: formation, galaxies: high-redshift, cosmology: theory, cosmology: dark ages, reionization, first stars, methods: numerical} 

\begin{document}

\begin{abstract}
Observations indicate that high-redshift galaxies undergo episodic star formation bursts, driving strong outflows that expel gas and suppress accretion. We investigate the consequences for metal and dust content of galaxies at $z\geq\!5$ using our semi-analytical model, \ash. We track gas-phase and stellar metallicities ($Z_\text{g}, Z_\star$) and dust mass ($\Mdust$) in dark matter haloes spanning $\Mhalo = 10^6\text{-}10^{11} \Msun$, comparing continuous and bursty star formation scenarios - which reflect underlying assumptions of instantaneous and delayed feedback - and we allow for metallicity-dependent feedback efficiency. Delayed feedback induces oscillations in $Z_{\rm g}$ and $Z_\star$, with $Z_{\rm g}$ declining sharply at low stellar and halo masses; the mass scale of this decline increases toward lower redshift. Reionization introduces significant scatter in $Z_{\rm g}$, producing an upturn followed by rapid decline. Metallicity-dependent feedback moderates this decline at $z=7\text{-}10$, flattening the $Z_\text{g}$–mass relation to $\simeq 0.03$–$0.04\,\Zsun$. Dust production tracks $Z_\text{g}$ but is sensitive to burst history, causing delayed enrichment.
Our results show that burst-driven feedback decouples $Z_{\rm g}$ and $Z_\star$, imprints intrinsic scatter in mass–metallicity relations, and delays dust growth. These effects are strongest in low-mass halos ($\Mhalo\sim 10^7\,\Msun$), where shallow potentials amplify the impact of feedback. Our results are consistent with recent hydrodynamical and semi-analytical simulations and provide context for interpreting JWST (\emph{James Webb Space Telescope}) metallicity and dust measurements, highlighting the importance of episodic star formation in early galaxy chemical evolution.
\end{abstract}

\section{Introduction}
\label{sec:introduction}
Recent estimates of the star formation histories of high-redshift galaxies using the JWST (\emph{James Webb Space Telescope}) indicate that star formation proceeds in bursts (i.e. is rapidly varying) in the early Universe \citep[e.g.][]{Faisst2019,Looser2023,Strait2023}. Bursty star formation occurs when the effective equilibrium between the self-gravity of gas in a galaxy and stellar feedback following star formation cannot be maintained \citep[cf.][]{FaucherGiguere2018}. It is expected to be commonplace in lower-mass galaxies \citep[e.g.][]{Onorbe2015,Muratov2015,Sparre2017} and, importantly, in galaxies at high redshifts \citep[e.g.][]{Pallottini2023,Shen2023,Sun2023}. 

Previous work has shown that bursty star formation will produce naturally strong stellar feedback-driven outflows in high-$z$ galaxies, especially at lower masses \citep[e.g.][]{Furlanetto2022, Menon2024}. This results in periods of oscillating gas mass in the galaxy as gas is expelled from the interstellar medium (ISM) and cosmological gas accretion from the intergalactic medium (IGM) is suppressed. The massive stars that are the sources of feedback, via e.g. stellar winds and supernovae \citep[e.g.][]{Lamers1999}, are also the sources of metals that enrich the ISM  \citep[e.g.][]{Mannucci.etal.2010,Maiolino.etal.2019} as well as the primary dust production mechanism in the early Universe \citep[e.g.][]{Todini2001,Gall2018}. 

This prompts the important question of \textit{how bursty star formation influences the metal and dust content of high-redshift galaxies.} 
From a theoretical perspective, we expect that metals and dust will play a crucial role in galaxy formation as coolants \citep[e.g.][]{Cox.1969,Sutherland.1993,Ploeckinger.2020}, which plays a critical role in regulating a galaxy's star formation and associated feedback efficiencies. From an observational perspective, a galaxy's metallicity will influence what we can infer from its spectral energy distribution \citep[e.g.][]{Robotham.etal.2020,Vijayan.etal.2025}, including diagnostics such as star formation rate. The presence of dust will obscure star forming regions in individual galaxies and star forming galaxies \citep[e.g.][]{Adelberger.2000}; there are indications from ALMA (\emph{Atacama Large Millimeter/submillimeter Array}) and JWST data  that galaxies at $z\geq\,6$ harbour a diverse range of dust content \citep[e.g.][]{Matthee.etal.2019,Rodighiero.etal.2023,Barrufet.etal.2023}. 

Astrophysically, the heavy elements that drive the growth of metallicity are the product of nucleosynthesis in the cores of stars, supernovae, and stellar mergers \citep[cf.][]{Nomoto2013,kobayashi.2020,Arcones2022}, which subsequently enrich the ISM, circumgalactic medium (CGM), and IGM via stellar-driven winds and supernovae. The physics of dust formation and evolution is complex \citep[e.g.][]{McKee.1989,Dwek1998,Calura2025}, but its production is believed to be driven primarily by supernovae at redshifts $z\geq5$ \citep[e.g.][]{Todini2001,Gall2011,Gall2018,Lesniewska2019,Ferrara2022,Langeroodi2024}. Feedback that drives gas out of a galaxy will lower the gas-phase metallicity of the ISM, whereas fresh accretion of nearly pristine gas from the IGM will lower the ISM gas phase metallicity. Similarly, supernovae-driven shocks in the ISM destroy dust, while they will also be entrained in the metal-enriched gas that is expelled from galaxies via winds. As a result, we might expect to see differences with respect to star formation modes in which burstiness is suppressed. This is because galaxies can retain more gas and more metal-enriched gas and dust in their ISM, which should allow for the gas-metallicity and dust content to increase steadily with time. Recent work by \cite{Marszewski2025} using the \textsc{FIRE-2} simulations suggests that bursty star formation is necessary to reproduce the nearly constant mass-metallicity relation at $z\geq5$ -- strong outflows induced by bursty star formation act to reset the ISM in a galaxy, and so metal-enriched accretion onto the galaxy is offset by reduced metal-production efficiency in the ISM. As \cite{Liu2024} show, bursty star formation's impact on metallicity, predominantly in lower-mass galaxies, introduces a strongly mass-dependent scatter that helps to explain ALMA [CII] metal line intensity mapping measurements during the Epoch of Reionization.

Interestingly, we might expect the efficiency of feedback to be influenced by metallicity at early epochs. The first generation of stars, the zero-metallicity Population III, enriched the ISM of their host galaxies \citep{Sluder2015,Chen2024} as supernovae \citep{Heger2002,Heger2010}, which led to subsequent generations of Population II stars being metal enriched. We expect increased stellar metallicity to influence stellar mass loss rates via winds, and consequently the efficiency of stellar feedback \citep[e.g.][]{Krticka2014,GormazMatamala2022,Rickard2022,Dekel2023}. It also regulates the maximum progenitor mass above which stars collapse directly into black holes without exploding as supernova \citep[e.g.][]{Zhang2008,O’Connor2011,Jecmen2023}, which reduces the stellar feedback efficiency for a given stellar population. 

In this paper, we study how bursty star formation influences the metal and dust content of high-$z$ galaxies - at $z\geq 5$, corresponding to the Cosmic Dawn and the Epoch of Reionization - using our lightweight\footnote{Here we use `lightweight' because the model employs simplified semi-analytical equations rather than full hydrodynamical simulations, enabling rapid exploration of parameter space for reasonable physical models. \ash~takes $\sim$15 mins to process 100 haloes on a 2020 M1 Macbook Pro.} semi-analytical model \ash\footnote{\ash~is from Sanskrit for `horsemen' or `charioteers', and refers to the twin celestial healers in Hindu mythology who restore balance and vitality — mirroring the model's focus on the self-regulating interplay between star formation and feedback. \ash~is a development of the (then unnamed) model introduced in \citet[][]{Menon2024}}\citep[see][hereafter \citetalias{Menon2024}]{Menon2024}. We extend \ash{} to track the growth of gas-phase and stellar metallicities, and dust, and we also include a model to account for the metallicity-dependence of the efficiency of feedback. These additions allow us to explore how bursty star formation during reionization imprints scatter and asymmetry in mass–metallicity relations, and how feedback cycles delay chemical and dust enrichment in halos of different mass ranges. We do this by investigating how our results are sensitive to key model parameters, and comparing to a model in which feedback is instantaneous and star formation proceeds in a smooth (i.e. non-bursty) manner.

\par The paper is structured as follows. In \S\ref{ssec:models}, we briefly review \ash~as implemented in \citetalias{Menon2024} before providing a detailed description on our updates to metal (\S\ref{ssec:metals_model}) and dust (\S\ref{subsec:Dust_Mass}) evolution, and metallicity-dependent feedback (\S\ref{ssec:feedback_model}). In \S\ref{sec:results}, we present our main results, reviewing our predictions for gas-phase and stellar metallicity growth over cosmic time; the relationship between metallicity and stellar mass; and predictions for dust growth over time. We review these results in the context of previous observational and theoretical work in \S\ref{sec:discussion}, and summarise our conclusions in \S\ref{sec:conclusions}. 

\par Note that we use the following values for the cosmological parameters: $\Omega_\text{b}=0.0484$, $\Omega_\text{M}=0.308$, $\Omega_\Lambda=0.692$, $h=0.678$, $\sigma_8=0.815$, and $n_\text{s}=0.968$, which are consistent with the results obtained by \citet{Planck2018}. We assume a metal mass fraction in the solar neighbourhood of $\Zsun=0.015$ \citep{Asplund2009,Lodders2019}.

\section{Theoretical Model}
\label{ssec:models}

\begin{table*}
    \centering
    \begin{tabular}{|p{1.5cm}|p{2cm}|p{2.5cm}|p{10cm}|}
        \hline
        \textbf{Parameters} & \textbf{Fiducial Value} & \textbf{Equation Number} & \textbf{Description} \\ \hline\hline

        $\varepsilon_{\rm UV}$ &  $[0,1]$  & --- & UV background suppression factor\\

        $z_{\rm rei}$ & 7 & --- & Redshift of reionization\\

        \hline
        \multicolumn{4}{|l|}{\textit{Parameters for modelling Star Formation and Supernova Wind feedback}}\\
        \hline

        $\epsilon_\text{sf}$ & $0.015$ & Equations~\ref{eq:sfr} and \ref{eq:metals_stars} & Star formation efficiency\\

        $\tau_\text{sf}$ & --- & Equations~\ref{eq:sfr} and \ref{eq:metals_stars} & Star formation timescale\\
        
        $t^\text{d} (=t-t^\prime)$ & $0.015$ Gyr & Equations~\ref{eq:gas_mass} and \ref{eq:SNe_wind}& Feedback delay timescale\\
        
        $\eta_\text{fb}$ & --- & Equations~\ref{eq:SNe_wind} and \ref{eq:eta_fb} & Mass-loading factor\\

        $\epsilon_\text{fb}$ & $5$ & Equation~\ref{eq:eta_fb} & Fraction of momentum from supernova that drives feedback winds\\

        $\pi_\text{p}$ & $1$
        & Equation~\ref{eq:eta_fb} & Total momentum injected per supernova (in units of $2 \times 10^{33} \text{ g cm s}^{-2}$)\\
   
        \hline
        \multicolumn{4}{|l|}{\textit{Parameters for modelling Gas and Stellar Metallicity}}\\
        \hline
                
        $Y_Z$ & $0.06$ & Equation~\ref{eq:metals_gas} & Metal yield per unit star formation\\
        
        $Z_\text{IGM}$ & $10^{-3} \Zsun$ & Equation~\ref{eq:metals_gas} & Metallicity of accreted intergalactic gas\\

        \hline
        \multicolumn{4}{|l|}{\textit{Parameters for modelling dust}}\\
        \hline

        $Y_\text{d}$ & $0.004$ & Equation~\ref{eq:dust_mass} & Dust yield per unit star formation\\

        $\tau_\text{dest}$ & --- & Equation~\ref{eq:dust_dest} & Dust destruction timescale\\

        $R_{\mathrm{SN}}$ & --- & Equation~\ref{eq:SNe_rate} & Rate of supernovae that destroys ISM dust \\
        
        $\gamma$ & $1.077\times 10^{-2}$ & Equation~\ref{eq:SNe_rate} & Fraction of stars (within $8-40 \Msun$ mass range) resulting in core-collapse supernovae\\

        $\epsilon_{\mathrm{d}} M_\text{SN,sw}$ & $10^3 \Msun$ & Equation~\ref{eq:dust_SN} & Efficiency of dust destruction via ISM swept-up by supernovae\\
        
        $M_\text{crit}$ & $10^5\Msun$ & Equation~\ref{eq:dust_SN} & Critical mass\\

        \hline
    \end{tabular}
    
    \caption{Parameter values for our fiducial model}
    \label{tab:model_params}
\end{table*}

\label{sec:model}
We use our lightweight semi-analytical model \ash,\footnote{The model is made publicly available at \href{https://github.com/Anand-JM97/Ashvini}{https://github.com/Anand-JM97/Ashvini}.} introduced in 
\citetalias{Menon2024}, to study the evolution of gas and stellar components in dark matter halos, which is motivated by the equilibrium model approach of \cite{Dave2012}. The model allows for delays between when stars form and when they produce feedback \citep[e.g.][]{Furlanetto2022} and time-dependent suppression of cosmological gas accretion, which reflects the growth of an ionizing UV background (UVB) in the early Universe \citep[e.g.][]{Kravtsov2022}. Note that we do not account for the effects of the high-$z$ progenitors of super-massive black holes and the feedback they produce; this will be incorporated in a forthcoming update to \ash. 

\subsection{Input halo merger trees}
\label{ssec:input_trees}
We use the same merger trees as in 
\citetalias{Menon2024}, which were generated with the Monte Carlo algorithm of \citet{Parkinson2008}. Each merger tree consists of a halo mass assembly history in the redshift range $5\leq\!z\leq\!25$, equally spaced in the logarithm of the expansion factor, $a = 1/(1 + z)$. We consider 11 mass bins at $z=5$ that are spaced in a quasi-logarithmic progression, alternating between factors of 10 and 5, within the mass range $10^6\leq M_{\rm h}/\Msun \leq 10^{11}$. We generate merger trees for 100 halos in each mass bin, such that we have 100 halos with masses of e.g. $M_\text{h}=10^6\Msun$ at $z=5$. Each merger tree provides us with $M_\text{h}$ as a function of z, and so we can calculate $\dot{M}_\text{h}$ as a function of redshift ($z$) or time ($t\equiv\!t(z)$) by constructing a smooth spline interpolant, which we differentiate numerically. In \citetalias{Menon2024} we had shown that the mass bin $10^7 \Msun$ marks the transition from `low' to `high' mass halos. Low/high-mass halos are particularly susceptible/robust to feedback processes due to their shallow/deep potential wells.

\subsection[The Ashvini Model]{The \ash~model}
In this subsection, we expand on the new updates to the \ash\,model. First, we give a brief summary of the model; we refer the reader to \citetalias{Menon2024} for a more detailed discussion.

\subsubsection{Core Model}
\ash~employs a set of coupled differential equations to describe the evolution of gas and stellar mass within a dark matter halo across cosmic time, and enforces mass conservation by balancing inflows, outflows, and internal processes. At a given instant in time, $t$, we compute the growth rate of the gas mass $\dot{M}_\text{g}$, the star formation rate $\dot{M}_{\star}$, and the wind-driven mass loss rate $\dot{M}_{\rm w}$ from the following:

\begin{align}
\dot{M}_\text{g}&=\dot{M}_\text{c,g}-\dot{M}_{\star} - \dot{M}_\text{w}(t^\prime),\label{eq:gas_mass}\\
\dot{M}_{\star}&=\epsilon_\text{sf}\frac{M_{g}}{\tau_\text{sf}},\label{eq:sfr}\\
\dot{M}_\text{w}&=\eta_\text{fb}\dot{M}_\star(t^\prime)\label{eq:SNe_wind},
\end{align}
where we assume the $\dot{M}$ values correspond to the instantaneous rates at $t$. $t^\prime = t - t^{\rm d}$ allows for a time delay ($t^{\rm d}>0$) between when stars form and when they produce feedback;  $t^{\rm d}=0$ corresponds to the case of instantaneous feedback. In Equation~\ref{eq:gas_mass}$, \dot{M}_\text{c,g}$ is the cosmological gas accretion from the IGM (see below); in Equation~\ref{eq:sfr}, $\Mg$ is the instantaneous mass of the gas reservoir, and $\epsilon_\text{sf}$ and $\tau_\text{sf}$ are the star formation efficiency and the star formation timescale (see below) respectively; and in Equation~\ref{eq:SNe_wind}, $\eta_\text{fb}$ is the feedback efficiency.  

As in \citetalias{Menon2024}, we estimate the cosmological gas accretion rate using,
\begin{equation*}
\dot{M}_\text{c,g}=\varepsilon_{\rm UV}\bigg(\frac{\Omega_\text{b}}{\Omega_\text{M}}\bigg)\dot{M}_\text{h},
\end{equation*}
where $\dot{M}_\text{h}$ is the halo mass growth rate; $(\Omega_\text{b}/\Omega_\text{M})$ is the cosmic baryon fraction; and $0\leq\varepsilon_\text{UV}\leq1$ quantifies the degree to which the accretion rate is suppressed by the UVB, which we assume is effective from a reionization redshift, $z_\text{rei} (=7)$. The specific value for $\varepsilon_\text{UV}$ at a given $z$ is fixed by our adopted model for UVB suppression from \citetalias{Menon2024}, as summarised in \ref{appendix:epsilon_uv}. We also assume that $\tau_\text{sf}$ is a fraction of the Hubble time, 
\begin{equation*}
        \tau_\text{sf} = 0.15 \frac{f_\text{sf}}{H(z)}.\label{eq:sfr_timescale}
\end{equation*}
We set $f_\text{sf}$ = 1 for simplicity. In practice we expect $f_\text{sf}$<1 in local star forming regions where dynamical times will be much shorter than the halo-averaged value, but $f_\text{sf}$ provides a reasonable upper limit given the uncertainties.

\subsubsection{Modelling Gas and Star Metallicity}
\label{ssec:metals_model}
In this work, we update the \ash~model to track the growth of the mass of metals in gas and stars, $\Zg$ and $\Zstar$. We extend Equations~\ref{eq:gas_mass}-\ref{eq:SNe_wind} as: 
\begin{align}
\dot{M}_{\text{Z,g}}&=Z_\text{IGM}\dot{M}_\text{c,g} + Y_Z\dot{M}_\star(t^\prime)-\dot{M}_{\text{Z},\star}-\dot{M}_{\text{Z,w}}(t^\prime),\label{eq:metals_gas} \\
\dot{M}_{\text{Z},\star}&=\epsilon_\text{sf}\dfrac{M_\text{Z,g}}{\tau_\text{sf}},\label{eq:metals_stars}\\
\dot{M}_{\text{Z,w}}&=\frac{M_\text{Z,g}}{M_\text{g}}\dot{M}_\text{w},\label{eq:metals_wind}
\end{align}
where $Z_\text{IGM}$ is the metallicity of the freshly accreted IGM gas and $Y_{\rm Z}$ is the heavy element yield per unit star formation rate (SFR). 
\begin{itemize}
\item Equation~\ref{eq:metals_gas} tracks the metal content of the accreting gas, which we assume to be at a fixed value $Z_\text{IGM}$ times the cosmological mass accretion rate $\dot{M}_\text{c,g}$; the metals formed in the cores of stars that would eventually enrich the ISM once the star dies, $Y_Z\dot{M}_\star(t-t_\text{d})$; the metals in the ISM that are locked in newly formed stars, $\dot{M}_{\text{Z},\star}$; and the metals that are lost via winds $\dot{M}_{\text{Z,w}}$.  
\item Equation~\ref{eq:metals_stars} assumes that the mass of metals in stars tracks the star formation rate $\dot{M}_\ast$ as given by Equation~\ref{eq:sfr}, times $(\Zg/\Mg)$, at that instant of cosmic time.
\item Equation~\ref{eq:metals_wind} assumes that the gas mass of metals lost via winds is simply $(\Zg/\Mg)$ times $\dot{M}_\text{w}$ at that instant, which is driven by the star formation rate $\dot{M}_\ast$ at the earlier time, $t^\prime$.
\end{itemize}

\noindent These equations assume an instantaneous perfect mixing of metals with the ambient gas reservoir. This is a simplifying assumption, but it is consistent with the rapid mixing timescales expected in a turbulent, feedback-driven ISM \citep[][]{Pan.2013,Hirai.2017} that we would expect in high redshift galaxies.

We use as our fiducial values $Z_\text{IGM}=10^{-3}\Zsun$, following \citet{Kravtsov2022}, and $Y_Z=0.06$, following \citet{Vincenzo2016} assuming a \citet{Chabrier2003} initial mass function (IMF). 

\subsubsection{Modelling Metal-Dependent Feedback}
\label{ssec:feedback_model}
We employ a momentum-regulated feedback recipe that balances the momentum injected into the ISM gas, by supernovae as well as other feedback processes, to that which is required to eject the gas at the halo's escape velocity \citep[also see][]{Furlanetto2017, Furlanetto2022}. This yields a prescription that depends on the redshift, $z$, and the halo mass, $M_\text{h}$ as, 
\begin{equation}    
\eta_\textbf{fb}=\epsilon_\text{fb}\pi_{\rm p}f(Z_\star)\bigg(\frac{10^{11.5}\Msun}{M_\text{h}}\bigg)^{1/3}\bigg(\frac{9}{1+z}\bigg)^{1/2}.\label{eq:eta_fb}
\end{equation}
Here $\pi_{\rm p}$ (which we set to unity) is the total momentum per supernova \citep[in units of $2\times10^{33}$ g cm s$^{-2}$, see also][]{Furlanetto2017, Furlanetto2022, Menon2024} and $\epsilon_{\rm fb}$ is the fraction of this momentum that is coupled to the galaxy's ISM as a wind.

The factor $f(Z_\ast)\leq1$ governs how the feedback efficiency depends on the stellar metallicity, $Z_\ast$. If we assume that there is no metal dependence, then $f(Z_\ast)=1$. However, we expect that the feedback should be metal-dependent at low stellar metallicities \citep[eg.][]{Zhang2008,O’Connor2011,Sukhbold2016,Jecmen2023}. In our model, we adopt an empirical parameterisation for $f(Z_\star)$ motivated by the work of \citet{Jecmen2023}. They estimated reductions in the total integrated momentum and mechanical energy of 75\% and 40\% respectively for stellar metallicities $Z_\star\lesssim 0.4 \Zsun$ compared to values expected at solar metallicity. We model this effect by means of a sigmoid function, which produces step-like behaviour with a smooth transition, given by,
\begin{equation}
    f(Z_\star)=\frac{\left(b-a\right)}{1+\exp-((Z_\star-m)/s)}+a. \label{eq:sigmoid_func}
\end{equation}
Here, $m=0.4\,\Zsun,~s=0.0067\,\Zsun,~a=0.25$ and $b=1$ are parameters that control the amplitude and sharpness of the transition of the sigmoid; this gives $f(Z_\star)=0.25$ at zero-metallicity. 

\subsubsection{Modelling Dust}
\label{subsec:Dust_Mass}
Although the physics of dust formation and evolution is complex \citep[e.g.][]{McKee.1989,Dwek1998,Calura2025}, we incorporate an approximate model to obtain estimates of the dust mass associated with the gas phase, $\Mdust$. At redshifts $z\geq5$, dust production is believed to be driven primarily by supernovae \citep[e.g.][]{Todini2001,Gall2011,Gall2018,Lesniewska2019,Ferrara2022,Langeroodi2024} because channels linked to evolved lower mass stars (e.g. the asymptotic giant branch, AGB) are unimportant until later cosmic times \citep{Marassi2019,Triani2020}. The standard assumption is that dust is produced by core-collapse supernovae of stars in the mass range $8-40\,\Msun$ \citep[cf.][]{Heger2003, Gall2018} and the dust yield per supernova is a function of the metallicity and mass of the progenitor mass. \cite{Gall2018} estimate this dust yield per supernova as $0.31\Msun$ with an uncertainty of 0.15 dex.
Higher redshift galaxies experience more efficient dust removal than their lower redshift counterparts, driven by active galactic nucleus (AGN) activity, supernova shocks, and astration, with recent estimates of dust removal rates showing a decline from 1.8 Gyrs at $z\simeq0.05$ to 450 Myrs at $z\geq3$ \citep[cf.][]{Lesniewska2025}.

In \ash, we assume that the rate of dust mass growth tracks the rate of core-collapse supernovae, which is proportional to $\dot{M}_\star$, and is lost through destruction in supernova shocks $(\dot{M}_\text{d,dest})$ and expulsion via winds $(\dot{M}_\text{w})$. We describe this using the equations,  
\begin{equation}
\dot{M}_\text{d}=Y_\text{d}\dot{M}_\star(t^\prime)- \dot{M}_\text{d,dest}(t^\prime)-\dot{M}_\text{d,w}(t^\prime),\label{eq:dust_mass}
\end{equation}
with 
\begin{equation}
\dot{M}_{\text{d,w}}=\left(\frac{\Mdust}{\Mg}\right)\dot{M}_\text{w}.\label{eq:dust_wind}
\end{equation}
Here, the total dust yield per unit star formation rate is $Y_\text{d}=0.004$ for our assumed \citet{Chabrier2003} IMF, following \citet{Gall2018} and \citet{Langeroodi2024}. The prefactor $(\Mdust/\Mg)$ in equation \ref{eq:dust_wind} assumes that dust mass loss via winds is proportional to the gas mass loss via winds - we assume an instantaneous perfect mixing of dust with the ambient gas reservoir - and $\dot{M}_\text{d,dest}$ determines the rate at which dust is destroyed by supernovae shocks.

The most straightforward parameterisation for $\dot{M}_\text{d,dest}$ is, 
\begin{equation}
\dot{M}_\text{d,dest}=\frac{\Mdust}{\tau_\text{dest}},\label{eq:dust_dest}
\end{equation}
where $\tau_\text{dest}$ is the dust destruction timescale \citep[e.g.][]{Triani2020}. While Equation~\ref{eq:dust_dest} is the conventional form, we instead adopt a parameterisation that explicitly links to the physical quantities important for dust destruction in our model. Because we assume that supernovae shocks drive dust destruction, we make explicit the dependence on the supernovae rate, which will govern the dust destruction rate, and choose, 
\begin{equation}
   \dot{M}_\text{d,dest}= \epsilon_\text{d}\,M_\text{SN,sw}R_\text{SN}\,f(\Mg/M_\text{crit}). \label{eq:dust_SN}
\end{equation}
Here $R_\text{SN}$ is the rate of supernovae, $M_\text{SN,sw}$ is the mass of the ISM swept up by supernovae ejecta, and $0\leq\epsilon_\text{d}\leq1$ governs the mass of dust in the swept-up ISM that is destroyed. In practice, we fix the product $\epsilon_\text{d}M_\text{SN,sw}$, as we show below. The function $f(\Mg/M_\text{crit})$ allows for supernovae to either sweep out dust from the potential in lower-mass galaxies or destroy dust in the ISM in higher-mass galaxies; $M_\text{crit}$ controls the transition between lower- and higher-mass galaxies.

\smallskip
\noindent\textbf{Estimating $R_\text{SN}$: }We evaluate $R_\text{SN}$ in our model at time $t$ using the star formation rate ($\dot{M}_\star$) and the number of core collapse supernovae per unit stellar mass formed, $\gamma$. $\dot{M}_\star$ is estimated at $t^\prime$ to account for the delay between when stars form and when they produce feedback. $\gamma$ can be calculated directly from the chosen IMF ($\xi$) as,
\begin{equation}
    \gamma = \frac{\int_8^{40}\xi(m)dm}{\int_{0.1}^{100}m\xi(m)dm}.
\end{equation}
Taken together, this gives,
\begin{equation}
R_\text{SN}=\gamma{\dot{M}_\star(t^\prime)},\label{eq:SNe_rate}
\end{equation}
where $\gamma$ varies between $1.014\times10^{-2}\Msun^{-1}$ for a \citet{Kroupa2001} IMF, to $1.077\times10^{-2}\Msun^{-1}$  for a \citet{Chabrier2003} IMF, to $1.958\times10^{-2}\Msun^{-1}$ for a \citet{Salpeter1955} IMF. We assume the \citet{Chabrier2003} as our default value in this paper.

\smallskip

\noindent\textbf{Estimating $\epsilon_\text{d}M_\text{SN,sw}$:} Previous work has estimated that $M_\text{SN,sw}$ is of order the amount of swept-up mass in the warm neutral medium (WNM) during the initial phase of a supernova, when the ejecta velocity is $\sim 10^4$ km s$^{-1}$, until it enters the Sedov phase. The WNM has temperatures and number densities $T\simeq10^4\text{K}$ and $n\sim0.1\text{cm}^{-3}$, and dust destruction occurs as a result of collisions with high-velocity ions ($\geq 100$ km s$^{-1}$). For a multiphase ISM, $M_\text{SN,sw}\lesssim10^4\Msun$ while $\epsilon_\text{d}=0.2$ \citep[e.g.][]{McKee.1989,Calura2025}. For simplicity, we assume that $\epsilon_\text{d}\,M_\text{SN,sw}=10^3\,\Msun$ \citep[e.g.][]{Gall2018,Calura2025}; the precise value
affects the normalisation but not the qualitative behaviour of dust evolution.
\smallskip

\noindent\textbf{Estimating $f(\Mg/M_\text{crit})$:} Dust destruction depends on whether the galaxy retains sufficient gas mass for supernovae to efficiently couple their energy to the ISM. We use the total gas mass $\Mg$ as a threshold: in low-mass systems ($\Mg<M_\text{crit}\simeq10^5\Msun$), supernova remnants break out of the galaxy before entering the Sedov-Taylor phase, ejecting dust via winds rather than destroying it in situ. We set $f(\Mg/M_\text{crit})\rightarrow0$ in this regime. In high-mass systems ($\Mg\geq M_\text{crit}$), remnants sweep up significant mass, efficiently destroying dust via shocks, and $f(\Mg/M_\text{crit})\rightarrow1$. To capture the smooth transition between these regimes, we adopt:
\begin{equation} 
f(\Mg/M_\text{crit}) = 1 - e^{-\,\left(\Mg/{M_\text{crit}}\right)^\alpha}, \label{eq:f_dust}
\end{equation} 
with $\alpha=8$ producing a sharp transition over $\sim\!1$ dex in mass. This allows for a range over which supernovae effects on dust shift from ejective to destructive. Dust mass in our model grows in proportion to the star formation rate (via core collapse supernovae), is destroyed by supernovae shocks in sufficiently massive galaxies, and is lost via winds in low-mass systems.

\section{Results}
\label{sec:results}

In this section, we present results for the trends in (gas and stellar) metallicities and dust mass with the \ash~model. We also explore the impact of different free parameters in our models on these trends. Table \ref{tab:model_params} lists our model parameters along with their fiducial values.

\subsection[Evolution of Mzg and Mzstar]{Evolution of $\Zg$ and $\Zstar$}

In this subsection, we focus on model predictions for the mass bin $M_\text{h}=10^7\Msun$ at $z=5$. We do this by tracking the behaviour of the median and $10^\text{th}$-to-$90^\text{th}$ percentile variation for gas and stellar mass, $\Mg$ and $\Mstar$, and mass of metals in gas and stars, $\Zg$ and $\Zstar$, as a function of cosmic time, using the Monte Carlo trees described in \S~\ref{ssec:input_trees}.
We assume the fiducial model parameters as given in Table~\ref{tab:model_params}. In those cases in which we show model predictions based on parameter variations (e.g. IGM metallicity, $Z_\text{IGM}$), we show the predictions of the fiducial model as light-grey curves. As argued in \citetalias{Menon2024}, $M_\text{h}=10^7\Msun$ is an interesting mass bin to examine because it marks the transition between low-mass halos that are quenched by delayed feedback alone and high-mass halos that continue to accrete gas and form stars to later times.

\medskip

\noindent\textbf{Fiducial Model Predictions:} We begin by assessing how the growth of $\Zg$ and $\Zstar$ is influenced by delayed feedback and the resulting bursty star formation. We compare our model predictions for the cases of instantaneous and delayed feedback, with the UV suppression of gas accretion in both cases. 

\begin{figure*}
\centering
\includegraphics[]
{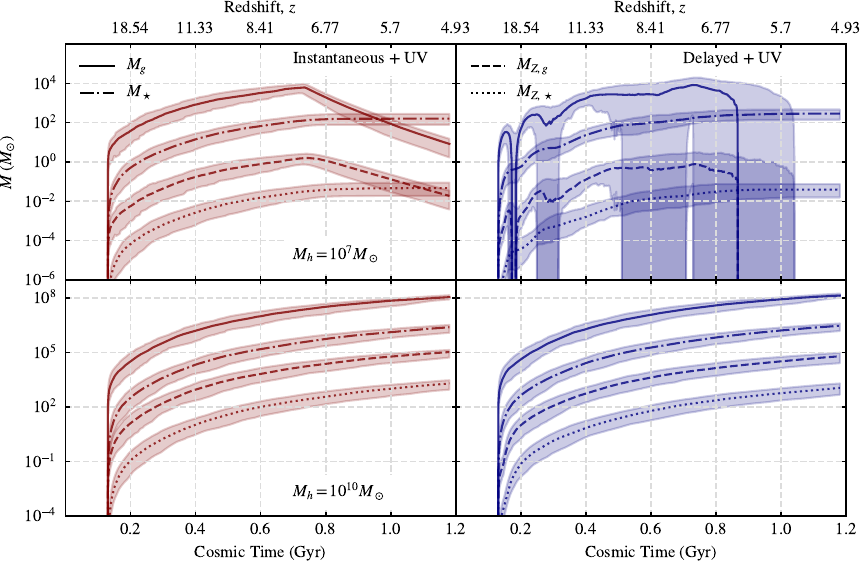}	
\caption{\textbf{Impact of instantaneous versus delayed feedback (fiducial model parameters):} The evolution of gas and stellar mass ($\Mg$ and $\Mstar$; solid and dotted-dashed curves), and the mass of metals in the gas phase and stars ($\Zg$ and $\Zstar$; dashed and dotted curves) as a function of cosmic time (in Gyrs, lower horizontal axis) and redshift (upper horizontal axis). These predictions are based on the assembly histories of 100 halos with $M_\text{h} = 10^7 \Msun$ ($M_\text{h} = 10^{10} \Msun$) mass bin at $z=5$ in the upper (lower) panel, for instantaneous (delayed) feedback in the left (right) panel. Each curve represents the evolution of the median value of a given quantity, while bands indicate the range of the 10$^{\rm th}$ and 90$^{\rm th}$ percentiles.}
\label{fig:FiducialGasStarMetals10e7_Inst_vs_Del}  
\end{figure*}

Figure~\ref{fig:FiducialGasStarMetals10e7_Inst_vs_Del} shows our model predictions assuming fiducial parameters for the instantaneous (delayed) feedback scenario (left and right panels) for halo mass bins of $M_\text{h} = 10^7 \Msun$ and $10^{10} \Msun$ (upper and lower panels). The solid and dot-dashed curves correspond to the median values of $\Mg$ and $\Mstar$; dashed and dotted curves correspond to the median values of $\Zg$ and $\Zstar$; and the shaded regions indicate the $10^\text{th}$-to-$90^\text{th}$ percentile variations in each quantity.

We first focus on the behaviour in the $M_\text{h} = 10^7 \Msun$ mass bin (upper panels). The case of instantaneous feedback shows that the evolution of $\Zg$ and $\Zstar$ closely follows that of $\Mg$ and $\Mstar$, respectively, with similar shapes but offset in normalisation. Both $\Mg$ and $\Zg$ show a smooth monotonic increase to their maximum value at $z\simeq7$, which corresponds to the redshift of reionization, $z_\text{rei}$, before a subsequent decline of $\sim\!3$ dex by $z\simeq 5$. $\Mstar$ and $\Zstar$ show a steady increase with cosmic time. The ratio of the two sets of curves remains consistent at $\Zg/\Mg=\Zstar/\Mstar\simeq10^{-4}$.

The case of delayed feedback shows some interesting differences with respect to the trends in the instantaneous feedback case. Both $\Mg$ and $\Zg$ show oscillatory behaviour at early times that closely track each other, although the oscillations are more pronounced in $\Mg$. The median $\Mg$ and $\Zg$ fall sharply to zero at $z\simeq 6.5$, although the $10^\text{th}$-to-$90^\text{th}$ percentile variations reveal that the metal-enriched gas persists in a subset of the halo sample to $z\simeq5.5$. There are imprints of the oscillations in $\Mg$ evident in $\Mstar$ and $\Zstar$ at early times, but otherwise they show a steady increase in cosmic time. Despite these differences, we see that the ratio of the two sets of curves mirrors that in the instantaneous case and remains consistent at $\Zg/\Mg=\Zstar/\Mstar\simeq10^{-4}$ while $\Mg>0$.

For comparison, the behaviour in the $M_\text{h} = 10^{10} \Msun$ mass bin (lower panels) reveals that the mode of star formation -- instantaneous or bursty -- has little-to-no effect on the growth of $\Mg$, $\Zg$, $\Mstar$ or $\Zstar$. Given this lack of sensitivity to star formation mode at higher masses, we focus on the $M_\text{h} = 10^7 \Msun$ mass bin in the remainder of this subsection.

Having established the fiducial behaviour of the instantaneous and delayed feedback scenarios, for the rest of this work, we limit our focus to only the delayed feedback scenario (but see \S\ref{sec:DustMass}).

\medskip

\noindent\textbf{Sensitivity to IGM metallicity, $Z_\text{IGM}$:}
\begin{figure}[ht!]
\centering
\includegraphics[width=\linewidth]{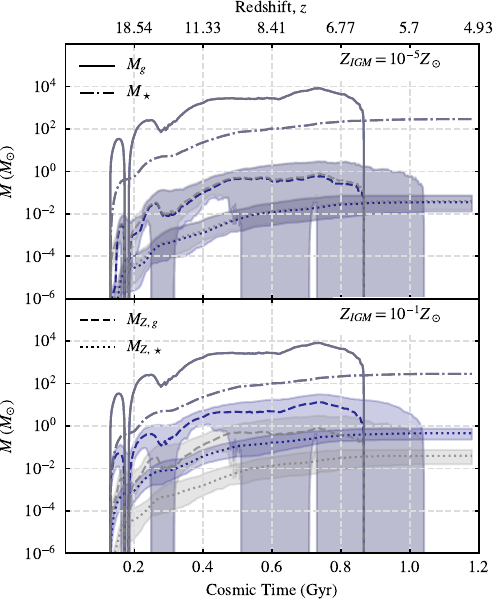}
\caption{\textbf{Influence of the IGM metallicity, $Z_\text{IGM}$:} We show the evolution of $\Mg$, $\Mstar$, $\Zg$, and $\Zstar$ (solid, dot-dashed, dashed, dotted curves, respectively) with cosmic time/redshift as the metallicity of the accreted gas $(Z_{\rm IGM})$ is varied. The upper and lower panels, respectively, corresponds to $Z_\text{IGM}=10^{-5}\Zsun$ and $10^{-1}\Zsun$. Grey bands and curves correspond to $\Zg$ and $\Zstar$ for the fiducial $Z_\text{IGM}$.}
\label{fig:FiducialGasStarMetals10e7_Inst_vs_Del_ZIGM}  
\end{figure}
Figure~\ref{fig:FiducialGasStarMetals10e7_Inst_vs_Del_ZIGM} shows how our model predictions depend on what we assume for the metallicity of IGM, $Z_\text{IGM}$. This sets the metallicity of the cosmologically accreted gas (cf. the first term in Equation~\ref{eq:metals_gas}). The upper (lower) panel shows the case for $Z_\text{IGM}$=$10^{-5}\,\Zsun$ ($10^{-1}\,\Zsun$); recall that the fiducial value is $10^{-3}\,\Zsun$ (grey bands and curves correspond to $\Zg$ and $\Zstar$ for this fiducial case). Note that we do not expect the value of $Z_\text{IGM}$ to affect the values of $\Mg$ and $\Mstar$ in any way and hence we do not show the interdecile ranges for these quantities.

For $Z_\text{IGM}=10^{-5}\,\Zsun$ (upper panels), the differences with respect to the fiducial predictions are negligible except at early cosmic times -- $\Zg$ and $\Zstar$ are marginally lower than in the fiducial case by less than 0.1 dex at $z\gtrsim 11$. For $Z_\text{IGM}=10^{-1}\,\Zsun$ (lower panels), differences with respect to fiducial predictions are more readily apparent -- both $\Zg$ and $\Zstar$ track the fiducial case but are offset by approximately 1 dex.

\medskip

\noindent\textbf{Sensitivity to  heavy element yield, $Y_Z$:}
\begin{figure}[ht!]
\centering
\includegraphics[width=\linewidth]{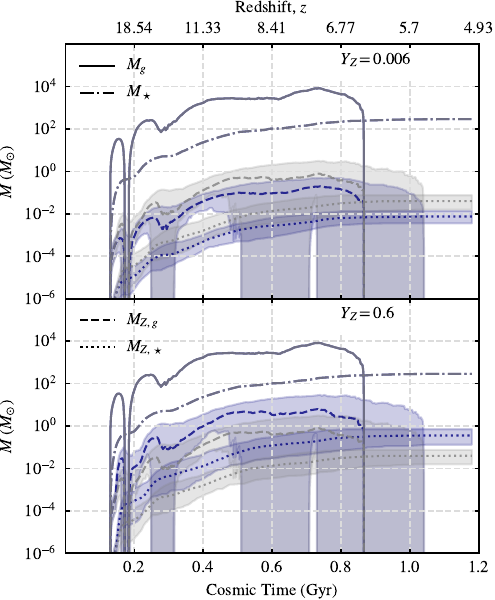}
\caption{\textbf{Influence of the heavy element yield, $Y_Z$:} We show the evolution of $\Mg$, $\Mstar$, $\Zg$, and $\Zstar$ (solid, dot-dashed, dashed, dotted curves, respectively) with cosmic time/redshift as heavy elements' yield $(Y_Z)$ is varied. Upper and lower panels corresponds to $Y_Z=0.006$ and $Y_Z=0.6$. Grey bands and curves correspond to $\Zg$ and $\Zstar$ for the fiducial $Y_Z$.\label{fig:FiducialGasStarMetals10e7_Inst_vs_Del_YZ}}
\end{figure}
Figure~\ref{fig:FiducialGasStarMetals10e7_Inst_vs_Del_YZ} shows how our model predictions are affected by what we assume for heavy element yield, $Y_Z$. This tracks the production of metals by (primarily massive) stars, which proceed to enrich the gas at the end of their main sequence life and are released into the galaxy's gaseous reservoir. The upper (lower) panel shows the case for $Y_Z$=0.006 (0.6); recall that the fiducial value is 0.06. As in Figure~\ref{fig:FiducialGasStarMetals10e7_Inst_vs_Del_ZIGM}, grey bands and curves correspond to $\Zg$ and $\Zstar$ for this fiducial value of $Y_Z$. As for variations in $Z_\text{IGM}$, we do not expect the value of $Y_Z$ to affect the values of $\Mg$ and $\Mstar$.

For $Y_Z=0.006$ (upper panel) and $Y_Z=0.6$ (lower panel), the differences with respect to the fiducial predictions are straightforward -- a decreased (increased) heavy element yield results in decreased (increased) values of $\Zg$ and $\Zstar$. The evolution of these quantities follows that in the fiducial case, with similar shapes but offset by $\sim\!1$ dex below (above) the fiducial curves for $Y_Z=0.006$ (0.6).

\medskip

\begin{figure}[ht!]
\centering
\includegraphics[width=\linewidth]{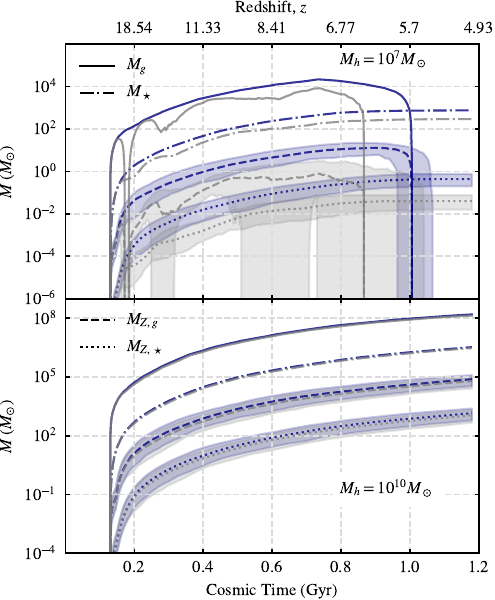}
\caption{\textbf{Influence of metal-dependent feedback:}  We show the impact of metal-dependent feedback (Equation~\ref{eq:sigmoid_func}) on the evolution of $\Mg$, $\Mstar$, $\Zg$, and $\Zstar$ (solid, dot-dashed, dashed, dotted curves, respectively) for lower and higher halo masses at $z=5$. The upper and lower panels correspond to $M_\text{h}=10^7\Msun$ and $10^{10}\Msun$ respectively. As before, the grey curves correspond to the fiducial model.}
\label{fig:SigmoidGasStarMetals_Inst_vs_Del}  
\end{figure}

\noindent\textbf{Metal-dependent feedback:} As noted earlier, there are good physical reasons to expect that the efficiency of feedback should be metal-dependent --- for example, because higher mass stars collapse directly into black holes \citep[e.g.][]{O’Connor2011,Sukhbold2016,Jecmen2023} or because the absence of metals in their outer envelopes reduces the mass and momentum flux of winds \citep[cf.][]{Lamers1999}. Figure~\ref{fig:SigmoidGasStarMetals_Inst_vs_Del} shows the impact of our assumed form for metal-dependent feedback (cf. \S~\ref{ssec:feedback_model}) on our model predictions for lower and higher halo masses - $M_\text{h}=10^7\Msun$ (upper panel) and $10^{10}\Msun$ (lower panel). Equation~\ref{eq:sigmoid_func} for $f(Z_\star)$ implies that there is a decrease in total momentum at sub-solar metallicities, and therefore a decrease in feedback efficiency at lower metallicities. 

The effect of the reduced feedback efficiency is readily evident for the case of $M_\text{h}=10^7\Msun$: there is stronger growth in the gaseous and stellar components when the feedback efficiency is metal-dependent: $\Mstar$ is offset by $\sim\!0.5$ dex, $\Zstar$ by $\sim\!1$ dex. 

The differences between $\Mg$ and $\Zg$, relative to the fiducial case, are more marked and qualitative. 
The weaker feedback at low metallicities and early times acts to erase the oscillations that are so apparent in the fiducial case. $\Mg$ has an amplitude similar to that in the fiducial case, but the median system can retain the gas mass for $\sim\!0.1$Gyr longer. The same behaviour is evident in the median value of $\Zg$, while the large fluctuations in the 10$^\text{th}$-to-90$^\text{th}$ variation evident in the fiducial case are eliminated.

This contrasts with what we observe at the higher halo mass of $10^{10}\Msun$, in the lower panel. 
At these masses, it is evident that the impact of metal-dependent feedback is negligible. The feedback mechanism plays a secondary role compared to the dominant influence of the deep gravitational potential of these haloes.

\subsection{Mass-metallicity relations}

\begin{figure*}[ht!]
\centering
\includegraphics[width=0.75\linewidth]{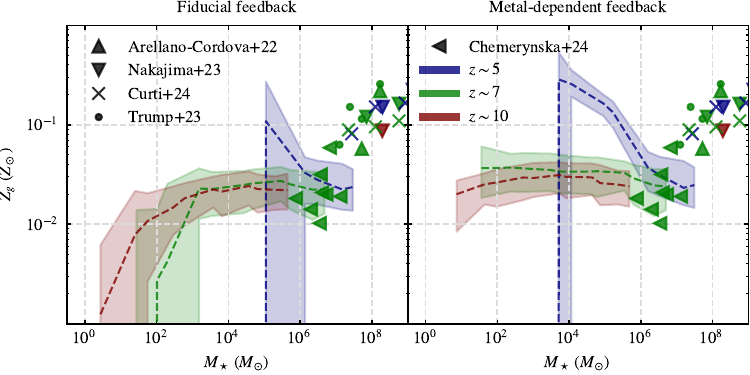}
\caption{\textbf{Stellar mass versus gas-phase metallicity relations between $z=5$ and $10$:} Here we plot gas phase  metallicity, $Z_{\rm g}=\Zg/\Mg$, as a function of stellar mass, $\Mstar$, in units of $\Msun$ at $z=5$, 7, and 10 (blue, green and red curves respectively). The left panel shows the behaviour in our fiducial feedback model; the right shows the impact of our assumed metal-dependent feedback model. Filled symbols correspond to observational data from \citet{ArellanoCordova2022}; \citet{Nakajima2023}; \citet{Trump2023}; \citet{Chemerynska.etal.2024}; and \citet{Curti2024}.}
\label{fig:StellarMassMetallicityRelation_gasphase}
\end{figure*}
\begin{figure*}[ht!]
\centering

\includegraphics[width=0.75\linewidth]{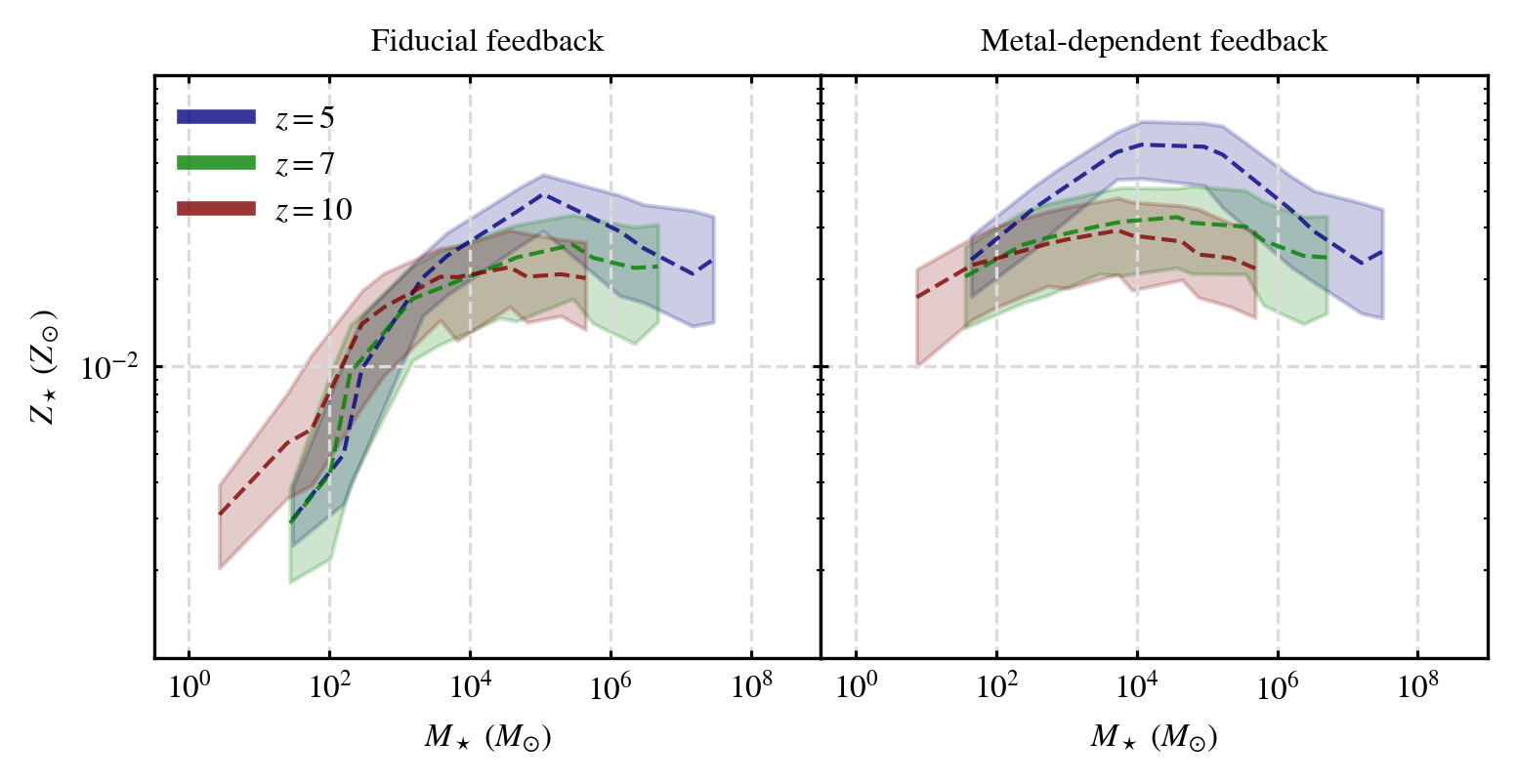}

\caption{\textbf{Stellar mass versus stellar metallicity relations between $z=5$ and $10$:} We show the same information as in Figure~\ref{fig:StellarMassMetallicityRelation_gasphase}, but for stellar metallicity, $Z_\ast=\Zstar/\Mstar$,
rather than gas-phase metallicity.}
\label{fig:StellarMassMetallicityRelation_stars}
\end{figure*}

In this subsection, we turn our attention to our model predictions for the relationship between stellar mass and the median gas-phase ($Z_g=\Zg/\Mg$) and stellar metallicities ($Z_\ast=\Zstar/ \Mstar$) at different epochs - Figures~\ref{fig:StellarMassMetallicityRelation_gasphase} and~\ref{fig:StellarMassMetallicityRelation_stars} respectively. As in the previous subsection, we focus on results for the delayed feedback scenario. 

\noindent\textbf{Gas-phase metallicity:} 
In Figure~\ref{fig:StellarMassMetallicityRelation_gasphase}, we show how gas-phase metallicity, $Z_{\rm g}$, varies with stellar mass, $\Mstar$, at redshifts $z=5$, 7, and 10 (blue, green and red curves respectively), for our fiducial feedback model (left panel) and metal-dependent feedback (right panel). The curves represent the median behaviour for a given $\Mstar$ at that epoch; the associated coloured bands indicate the $10^\text{th}$ to $90^\text{th}$ percentile variation. The symbols correspond to observational data from \citet{ArellanoCordova2022}, \citet{Nakajima2023}, \citet{Trump2023}, \citet{Chemerynska.etal.2024}, and \citet{Curti2024}.

The fiducial model predicts that the metallicity at higher stellar mass ($\Mstar$) is relatively flat at a given epoch, with $Z_{\rm g} \simeq 0.03\,\Zsun$. However, at lower masses, the metallicity declines sharply, with a corresponding increase in the scatter in $Z_{\rm g}$. The mass scale at which this decline occurs increases with decreasing redshift. At $z=7$ and $z=10$, there is a monotonic decline in $Z_{\rm g}$. However, at $z=5$, we observe a sharp upturn followed by a precipitous drop. This behaviour at $z=5$ is an artifact of the onset of UV suppression in $z_\text{rei}=7$, which introduces a large scatter in the values of $\Mg$ and $\Zg$. This scatter is evident in Figure~\ref{fig:FiducialGasStarMetals10e7_Inst_vs_Del}, particularly in the $10^\text{th}$ to $90^\text{th}$ percentile variation, and affects $Z_{\rm g}$ due to the small value ratios. Some of this scatter arises from stochastic halo growth histories in the merger trees, which, together with UV suppression, amplify the apparent upturn and subsequent drop.

In contrast, the effect of metal-dependent feedback is to suppress the decrease in $Z_{\rm g}$ at low $\Mstar$ and $M_{\rm h}$ at $z=7$ and $z=10$. The relation remains relatively flat, with $Z_{\rm g} \simeq 0.03$–$0.04\,\Zsun$ across the entire mass range. A decline in $Z_{\rm g}$ becomes evident only at $z=5$, showing the same sharp upturn and precipitous drop seen in the instantaneous case, but occurring at a stellar mass roughly a factor of $\sim\!10$ smaller.

\medskip 

The decline in gas-phase metallicity ($Z_{\rm g}$) with decreasing mass at $z = 10$ and $z = 7$ reflects the increasing efficiency of feedback-driven metal ejection in shallow potential wells. By $z = 5$, this trend reverses: low-mass halos exhibit higher $Z_{\rm g}$ because reionization-driven quenching has already expelled most of their gas, and metal-enriched ejecta from the most recently formed stars enrich the small remaining reservoirs. This process produces the observed upturn in our model. At even lower masses, stellar feedback combined with UV suppression of cosmological gas accretion is sufficient to remove gas entirely from these systems.

These predictions imply that multi-band photometry combined with metallicity measurements can discriminate between feedback scenarios. For example, delayed feedback generates larger scatter in the $Z_{\rm g}$–$\Mstar$ relation ($\Delta\log Z_{\rm g} \sim 0.5$-$1$ dex) and sharper declines at characteristic masses that evolve with redshift.


\medskip

\noindent\textbf{Stellar metallicity:} 
Figure~\ref{fig:StellarMassMetallicityRelation_stars} 
mirrors Figure~\ref{fig:StellarMassMetallicityRelation_gasphase}; here we show how stellar metallicity, $Z_{\star}$, varies with stellar mass, $\Mstar$ at different redshifts assuming our fiducial and metal-dependent feedback models (left and right panels respectively). 

We see that $Z_\star$ decreases with decreasing $\Mstar$
for all redshifts. There is strong overlap between the relations at different redshifts, as we might expect; $\Mstar$ and $\Zstar$ will either remain constant or grow in time, which contrasts with $\Mg$ and $\Zg$, which can either increase or decrease in response to cosmological gas accretion, enrichment episodes via supernovae, and expulsion via stellar-driven winds. 
Compared to the behaviour we see in $Z_{\rm g}$ with $M_\ast$,  there is a sharper decline in $Z_\ast$ with decreasing $M_\ast$
and an increase in the 10$^\text{th}$-90$^\text{th}$ percentile variation. 


\subsection[Evolution of Dust, Md]{Evolution of Dust, $M_{\rm d}$} \label{sec:DustMass}

\begin{figure}[ht!]
\centering
\includegraphics[width=\linewidth]{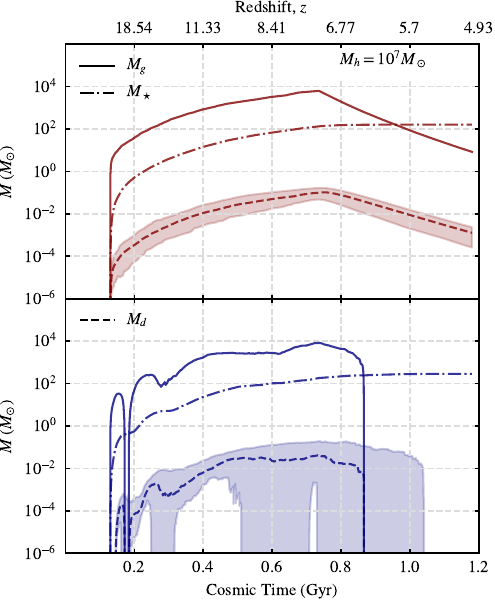}
\caption{\textbf{Evolution of dust mass $\Mdust$ for low-mass halos:} The growth of $\Mg$ (solid), $\Mstar$ (dash-dotted),  $\Mdust$ (dashed) in halos of mass bin (at $z=5$) $\Mhalo=10^{7}\,\Msun$. The upper(lower) panel is for the instantaneous(delayed) feedback scenario. For clarity we only show the median behaviour  for $\Mg$ and $\Mstar$ while the shaded region represents the $10^{\rm th}-90^{\rm th}$ variation in $\Mdust$.}
\label{fig:FiducialGasStarDust10e7_Inst_vs_Del}
\end{figure}

\begin{figure}[ht!]
\centering
\includegraphics[width=\linewidth]{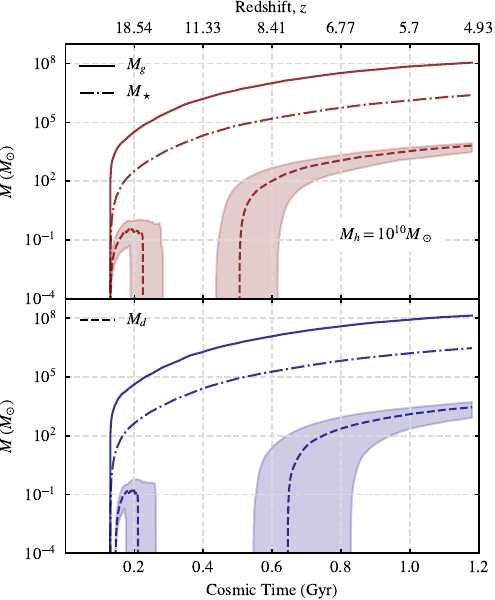}\caption{\textbf{Evolution of dust mass $\Mdust$ for high-mass halos:} Similar to Figure \ref{fig:FiducialGasStarDust10e7_Inst_vs_Del} we show the $\Mg$ (solid), $\Mstar$ (dash-dotted),  $\Mdust$ (dashed) but for halos of mass bin (at $z=5$) $\Mhalo=10^{10}\,\Msun$. 
}\label{fig:FiducialGasStarDust10e10_Inst_vs_Del}
\end{figure}

In this final subsection, we examine our model predictions for the dust mass, $\Mdust$. As usual, we focus on two representative systems at $z=5$: low-mass halos with $\Mhalo=10^7\,\Msun$ and high-mass halos with $\Mhalo=10^{10}\,\Msun$, and we consider the behaviour expected for both instantaneous and delayed feedback scenarios. Figures~\ref{fig:FiducialGasStarDust10e7_Inst_vs_Del} and~\ref{fig:FiducialGasStarDust10e10_Inst_vs_Del} show the evolution of gas, stellar, and dust masses (solid, dot-dashed, and dashed curves, respectively) as a function of cosmic time for these two halo masses. As in our other plots, we follow both the median trend and the $10^\text{th}$–$90^\text{th}$ percentile range of $\Mdust$.

Across both mass scales, $\Mdust$ broadly traces the behaviour of $\Mg$, offset in normalisation by a factor of $\sim\!1/100$, with a similar level of scatter. This is expected: our model links dust growth to the SFR, $\dot{M}_\ast$, which depends directly on $\Mg$. The evolution also mirrors that of the gas-phase metal mass $\Zg$ (see Figure~\ref{fig:FiducialGasStarMetals10e7_Inst_vs_Del}), reflecting the fact that, to the first order, $\Mdust$ can be obtained from $\Zg$ via a constant metal-to-dust ratio.  Consequently, the oscillatory behaviour of $\Zg$ is echoed in $\Mdust$. In our model, deviations from this constant ratio occur whenever dust destruction channels (equations \ref{eq:dust_dest}–\ref{eq:f_dust}) operate more efficiently than metal loss, particularly in high-mass halos where supernovae shocks dominate. The scatter in $\Mdust$ for a given SFR - which at fixed halo mass $\Mhalo$ will show a scatter as a result of variations in assembly history - implies that dust-based SFR estimates could be significantly biased if this intrinsic variation is not taken into account.

For low-mass $\Mhalo=10^7\,\Msun$ halos, $\Mg$ almost never exceeds $10^5\,\Msun$, so the primary pathway for dust removal is through stellar winds; these systems never reach the Sedov–Taylor phase of supernova evolution. In the instantaneous feedback scenario (upper panel of Figure~\ref{fig:FiducialGasStarDust10e7_Inst_vs_Del}), $\Mdust$ follows the gradual decline of $\Mg$ once the UVB is turned on at $z_{\rm rei}$. In contrast, in the delayed feedback scenario (lower panel), $\Mdust$ declines to zero along with $\Mg$, as the combined effects of the UVB and feedback rapidly reduce the gas reservoirs in these galaxies.

The high-mass $\Mhalo=10^{10}\,\Msun$ halos present a different picture. Their deeper gravitational potential wells allow them to retain more gas, enabling supernova shocks - in addition to stellar winds - to destroy dust. Once $\Mg$ is high enough for the Sedov–Taylor phase to develop, dust destruction becomes the dominant removal process, producing sharp declines in $\Mdust$. Only when $\Mg \gtrsim 10^7\,\Msun$ does dust production begin to balance destruction. As a result, our model predicts that high-mass systems can host ongoing star formation while exhibiting a wide diversity in dust content, from almost dust-free to dust-rich states. This diversity is qualitatively consistent with the oscillatory behaviour seen in more sophisticated simulations \citep[e.g.][]{Choban.etal.2025} and could reconcile observations of high-$z$ galaxies with a wide range of dust masses \citep[e.g.][]{Rodighiero.etal.2023,Barrufet.etal.2023}

\section{Discussion}
\label{sec:discussion}

We have demonstrated that bursty star formation fundamentally alters the chemical evolution of early galaxies compared to continuous star formation scenarios. The delayed feedback inherent to bursty star formation drives oscillatory behaviour in both gas-phase and stellar metallicities, while reionization introduces additional complexity by suppressing gas accretion. Together, these processes generate large scatter in gas mass ($\Mg$)) and gas-phase metallicity ($\Zg$), producing distinct signatures in the mass–metallicity relation that depend on the feedback prescription.

Our predictions align closely with recent JWST observations, particularly for galaxies with $\Mstar \sim 10^6\!-\!10^8\,M_\odot$. Figure~\ref{fig:StellarMassMetallicityRelation_gasphase} shows that delayed feedback combined with UV suppression reproduces the low normalization ($\Zg \sim 0.02\!-\!0.05\,Z_\odot$) and large intrinsic scatter ($\Delta \log \Zg \sim 0.5\!-\!1$ dex) observed at $z \simeq 5\!-\!10$. The steep decline in $\Zg$ at low masses and its evolution with redshift match the trends reported by \citet{Nakajima2023}, \citet{Curti2024}, and \citet{Chemerynska.etal.2024}, while the upturn and subsequent drop at $z \simeq 5$ reflect reionization-driven gas depletion inflating $\Zg$ in residual reservoirs before complete gas removal. Metal-dependent feedback flattens the relation at $z \simeq 7\!-\!10$, yielding $\Zg \sim 0.03\!-\!0.04\,Z_\odot$ across the mass range, consistent with the lower end of the observed normalization.

Observational metallicities from \citet{ArellanoCordova2022} and \citet{Sarkar2025} agree with the plateau and sharp decline predicted by our delayed feedback models at $z \simeq 6\!-\!10$. \citet{Trump2023} and \citet{Li2025} report higher values - 12+$\log(\mathrm{O/H}) \simeq 7.2\!-\!8.0$ ($\sim 0.03\!-\!0.18\,Z_\odot$) - that more closely resemble our instantaneous feedback + UV background results, highlighting the observational split between feedback scenarios. The lower normalisation of the mass–metallicity relation at $z \simeq 7\!-\!10$ noted by \citet{Heintz2022} and \citet{Ucci2022} is also reproduced, suggestive of pristine inflows and strong outflows in low-mass halos. Furthermore, the decoupling of gas-phase and stellar metallicities in our models mirrors episodic enrichment histories inferred by \citet{rey.2025} and elevated N/O ratios reported by \citet{ji.2025}.

Theoretical and simulation studies reinforce these trends. The \textsc{DELPHI23} semi-analytic model \citep{Mauerhofer2025} predicts a sharp metallicity drop at low masses, similar to our delayed + UV results at $z \simeq 5$. Cosmological hydrodynamical simulations such as \textsc{FIRE} and \textsc{FIRE-2} \citep{Ma2015,Hopkins2018} exhibit mass–metallicity relations whose shapes and scatter closely match our predictions for $\Mstar \sim 10^6\!-\!10^8\,\Msun$, including steep declines in $\Zg$ in burst-dominated regimes. In \textsc{FIRE}, the stellar metallicity–stellar mass relation also resembles our delayed + UV case, with offsets of only $\sim 0.5\!-\!2$ dex at comparable redshifts.

Our dust evolution predictions capture key features seen in recent high-$z$ surveys. The delayed growth and large scatter in dust mass ($\Mdust$) in our models parallel the diversity observed by \citet{Rodighiero.etal.2023} and \citet{Barrufet.etal.2023}, and match the quenching or decline in dust content seen in \textsc{FIRE-2} simulations \citep{Choban.etal.2025}. In contrast, the \textsc{CROC} simulations \citet{Esmerian2022} predict steady dust growth in massive halos, whereas our delayed feedback scenario produces a pronounced quench over similar timescales - likely reflecting our assumption that $M\_{\rm ISM} = \Mg$, which may overestimate dust destruction in gas-rich halos.

Taken together, these comparisons show that burst-driven, delayed feedback, when combined with reionization suppression, can naturally explain several observed high-$z$ trends - oscillatory enrichment, large scatter in $\Zg$, decoupled stellar and gas-phase metallicities, and stochastic dust evolution. At the same time, the split between observational cases favouring delayed versus instantaneous feedback underscores the need for joint constraints using metallicity, dust content, and star formation variability to fully pin down feedback timescales in early galaxies.

\medskip

\section{Conclusions}
\label{sec:conclusions}

We have investigated how bursty star formation during cosmic reionization affects the metallicity and dust content of low-mass galaxies using our \ash~model. Bursty star formation can drive strong outflows, which suppress accretion of pristine gas from the intergalactic medium. This can offset metallicity in these galaxies, and the episodic nature of enrichment can decouple gas-phase and stellar metallicities, which will influence how we interpret the observable properties of the high-redshift galaxy population. Our results confirm these expectations:
\begin{itemize}
\item \emph{Bursty star formation}, driven by delayed feedback, induces oscillations in gas and stellar metallicities ($Z_g$, $Z_\star$), particularly in lower-mass halos, as illustrated by our $\Mhalo \sim 10^7\,\Msun$ example (Figure~\ref{fig:FiducialGasStarMetals10e7_Inst_vs_Del}). This leads to a decoupling of $Z_g$ and $Z_\star$, consistent with the episodic accretion and enrichment histories inferred empirically by \citet{Chemerynska.etal.2024} using JWST data and theoretically by \citet{rey.2025} using the EDGE simulations \citep[Engineering Dwarfs at Galaxy Formation’s Edge; cf.][]{Agertz2020} of dwarf galaxies.

\item \emph{Reionization} introduces significant scatter in $Z_g$, producing a sharp upturn followed by a decline at $z \lesssim 7$ (Figure~\ref{fig:FiducialGasStarMetals10e7_Inst_vs_Del}, lower panel). This feature aligns with the metallicity plateau and subsequent drop observed in JWST studies \citep[e.g.][]{ArellanoCordova2022,Sarkar2025}.

\item \emph{Metal-dependent feedback} flattens the $Z_g$–$\Mstar$ relation to $Z_g \approx 0.03$–$0.04\,\Zsun$ at $z = 7$–10 (Figure~\ref{fig:StellarMassMetallicityRelation_gasphase}), in agreement with the lower normalization of the mass–metallicity relation reported by \citet{Heintz2022} and \citet{Ucci2022}.

\item \emph{Dust evolution} broadly tracks $Z_g$ but is delayed by feedback cycles. The resulting diversity in dust content, especially in high-mass halos, is consistent with ALMA and JWST observations of dust-rich and dust-poor galaxies at $z \gtrsim 6$ \citet{Rodighiero.etal.2023} and \citet{Barrufet.etal.2023}.
\end{itemize}

The results confirm that inferences of dust content or stellar age based on galaxy colour may be misleading in bursty, metal-poor systems. Because dust production follows gas-phase metallicity and is delayed by feedback, both the timing and amount of dust are sensitive to burst histories. Bursty star formation introduces time lags and variability in enrichment. The metallicity of gas inflow properties and the metallicity-dependence of supernovae can critically shape metallicity and dust evolution. These effects are especially strong in low-mass galaxies, where shallow potentials amplify burst effects. As such, observations of galaxies in the early Universe must account for these burst-driven dynamics when interpreting metallicity, dust, and star formation rates.


\section{Acknowledgement}
The authors thank the anonymous referee for their careful and thoughtful report. SB acknowledges the Dr Albert Shimmins Fund for a Postgraduate Writing-Up Award, and  support from grant PID2022-138855NB-C32 funded by MICIU/AEI/10.13039/501100011033 and ERDF/EU, and  project PPIT2024-31833, cofunded by EU-Ministerio de Hacienda y Función Pública-Fondos Europeos-Junta de Andalucía-Consejería de Universidad, Investigación e Innovación. CP acknowledges the support of the ARC Centre of Excellence for All Sky Astrophysics in 3 Dimensions (ASTRO 3D), through project number CE170100013. This research uses \textsc{python} \citep{vanRossum.1995}
libraries, including \textsc{numpy} \citep{Harris2020}, \textsc{matplotlib} \citep{Hunter2007}, and \textsc{scipy} \citep{Virtanen2020}.

\section*{Code \& Data availability}
The \ash~model - along with the dark matter halo merger trees used in this work - is made publicly available at 

\noindent\href{https://github.com/Anand-JM97/Ashvini}{https://github.com/Anand-JM97/Ashvini}.
\bibliography{ref}

\appendix
\section{Calculation of \texorpdfstring{$\varepsilon_\text{uv}$}{epsilon extunderscore uv}}
\label{appendix:epsilon_uv}
In \citetalias{Menon2024}, we used the parameter $\epsilon_\text{in}\equiv\,\varepsilon_\text{UV}$ to regulate cosmological gas accretion onto halos in the presence of a UVB, arguing that suppression of accretion will be dominated by the UVB at redshifts $z\gtrsim 5$. For completeness we summarise the arguments of \S~2.4.1 in \citetalias{Menon2024}, which builds on \citet{Gnedin2000}, \citet{Okamoto2008}, and \citet{Kravtsov2022}, to motivate the functional form of $\varepsilon_\text{UV}$.

\medskip

\noindent The baryon fraction within a halo is given by 
\begin{equation}
	f_\text{b}(M_\text{h},z)=\frac{\Omega_\text{b}}{\Omega_\text{m}}s(\mu_\text{c},\omega)
\end{equation}
where $\Omega_\text{b}/\Omega_\text{m}$ is the cosmic baryon fraction. $s(x,y)$ is defined as,
\begin{equation}
	s(x,y)=[1+(2^{\frac{y}{3}}-1)x^{-y}]^{-\frac{3}{y}}, 
\end{equation}
where $\mu_\text{c}=m_\text{h}/M_\text{c}(z)$. $M_\text{c}(z)$ is a characteristic mass below which $s(\mu_\text{c},\omega) \rightarrow 0$, 
\begin{equation}
	M_\text{c}=1.69\times 10^{10}\frac{\text{exp}(-0.63z)}{1+\text{exp}([z/\beta]^{\zeta})}M_{\odot},\label{2.2}
\end{equation}
where $\zeta=15$ and $\beta$ is given by,
\begin{equation}		\beta=z_\text{rei}\left[\text{ln}\Big(1.82\times10^3\text{exp}(-0.63z_\text{rei})-1)\right]^{-1/\zeta};
\end{equation}
here $z_\text{rei}$ is the redshift of reionization. 

\medskip

\noindent We write,
\begin{equation}                       
\epsilon_\text{in}=\text{max}\bigg(0,s(\mu_\text{c},\omega)\bigg[(1+X)- 2\epsilon(z,\zeta)\frac{M_\text{h}}{\dot{M}_\text{h}}X(1+z)H(z)\bigg]\bigg),\label{2.5}
\end{equation}
where 
\begin{align*}
\epsilon(z,\zeta)&=\frac{0.63}{1+e^{(z/\beta)^\zeta}}+\frac{\zeta z^{\zeta-1}}{\beta^\zeta}\frac{e^{(z/\beta)^\zeta}}{(1+e^{(z/\beta)^\zeta})^2},\\
X&=\frac{3c_\omega M_\omega}{1+c_\omega M_\omega},\quad c_\omega=2^{\omega/3}-1,\\ 
M_\omega&=\left(\frac{M_\text{c}(z)}{M_\text{h}}\right)^\omega,\quad\omega=2.
\end{align*}
The net effect is to suppress gas accretion at halo masses $M_\text{h}\leq\!M_\text{c}(z)$ at $z<z_\text{rei}$, with $M_\text{c}(z)$ increasing with decreasing $z$ (i.e. accretion is suppressed onto progressively higher mass halos). For redshifts $z>z_\text{rei}$, there is no suppression of gas accretion. 

\end{document}